Title: **Laser induced ultrafast Co 3d and Ho 4f spin dynamics in CoHo ferrimagnetic alloys.**

D. Gupta[1], B. Kunyangyuen[2], M. Riepp[2], K. Holldack[1], R. Abrudan[1], T. Kachel[1], N. Pontius[1], C. Schüßler-Langeheine[1], M. Vomir[3], M. Hehn[2] and N. Bergeard[3*]

[1] *Helmholtz-Zentrum Berlin für Materialien und Energie, Albert-Einstein-Str. 15, 12489 Berlin, Germany*

[2] *Institut Jean Lamour, Université de Lorraine, BP 50840, 54011 Nancy, France.*

[3] *Université de Strasbourg, CNRS, Institut de Physique et Chimie des Matériaux de Strasbourg, UMR 7504, F-67000 Strasbourg, France*

* Corresponding author:
*Mel:* nicolas.bergeard@ipcms.unistra.fr
*Address:* Institut de Physique et de Chimie des Matériaux de Strasbourg (IPCMS)
Campus Cronenbourg 23 rue du Loess BP43 67034 Strasbourg

## *0. Abstract*

The transition metal (TM) / rare-earth (RE) alloys have received renewed interest lately as model systems to decipher the origin of all-optical helicity-independent switching (AO-HIS) by femtosecond laser pulses. Recently, a distinct single-pulse magnetization reversal mechanism, based on domain-wall motion and coalescence, has been reported in CoDy and CoHo alloys, as well as in Co/Gd ultrathin bilayers. It has been claimed that this specific toggle switching is achieved when the Co sublattice is fully demagnetized and the characteristic demagnetization times $\tau_{RE}$ of the RE (Gd, Dy or Ho) sublattices is longer than that of the Co and the angular momentum transferred to Co is reduced. Element- and time-resolved X-ray spectroscopy studies of CoDy alloys have reported characteristic demagnetization times of $\tau_{Co} \approx 0.2$ ps and $\tau_{Dy} \approx 0.6$ ps at room temperature. Similarly, Ho is expected to exhibit a slower response, but its ultrafast 4f spin dynamics remain experimentally unexplored. Here, we report on element- and time-resolved investigations of femtosecond laser induced ultrafast dynamics of the Co 3d and Ho 4f spins in ferrimagnetic $Co_{80}Ho_{20}$ alloys to verify this prediction. We observed characteristic demagnetization times $\tau_{Co} = 0.22 \pm 0.01$ ps and $\tau_{Ho} = 0.87 \pm 0.15$ ps at room temperature.  These results show that Ho demagnetizes substantially more slowly than Co in $Co_{80}Ho_{20}$, supporting the proposed condition for domain-wall-mediated all-optical toggle switching.

*Highlights*:

- Element- and time-resolved investigation of laser induced ultrafast Co 3d and Ho 4f spin dynamics in ferrimagnetic CoHo amorphous alloy at room temperature.

- Characteristic demagnetization time $\tau_{Ho} = 0.87 \pm 0.15$ ps for the Ho and $\tau_{Co} = 0.2 \pm 0.01$ ps for the Co sub-lattices at T = 300 K.

## *1. Introduction*

The excitation of magnetic layers with infra-red (IR) femtosecond (fs) laser pulses leads to a quenching of the magnetic order on a sub-picosecond time scale [BEA96]. The main challenge in ***"femtomagnetism"*** consists in identifying the microscopic mechanisms responsible for the ultrafast angular momentum (AM) transfer from spin- to electron- and lattice-degrees of freedom. Particularly interesting for studying the laser-induced ultrafast spin dynamics is the class of ferrimagnetic rare-earth / transition-metal (RE-TM) alloys with their antiferromagnetically coupled RE 4f and TM 3d sublattices [BRO91]. Ultrafast spin dynamics in RE-TM alloys shows very different behaviour compared to the one reported in pure TM [KOO10] and pure RE earth layers [WIE11]. Furthermore, the laser-induced ultrafast TM 3d and RE 4f spin dynamics in these alloys are not concomitant [RAD11] and show distinct dependence on temperature [FER21]. In the latter study [FER21], authors showed that, in CoDy, with increase in temperature, the characteristic demagnetization time for Co ($\tau_{Co}$) remains nearly constant while it decreases for Dy ($\tau_{Dy}$). In contrast, in both pure TM and RE layers, the characteristic demagnetization time is known to increase with temperature [ROT12, SUL12].

Since the discovery of all-optical helicity-independent switching (AO-HIS) in FeCoGd [VAH09, RAD11], these alloys have received renewed interest. Two mechanisms for such magnetization reversal by single laser pulse were identified: the ultrafast exchange-mediated mechanism observed in Gd based materials such as FeCoGd alloys [STA06, RAD11] and Co/Gd bi-layers [LAL19] as well as a slower precessional mechanism observed in Co/Tb bi-layers [MIS23], TM/Tb bi-layers or CoRE/TM/CoRE trilayers [PEN23]. Kunyangyuen et al. have recently reported on a third mechanism based on domain wall motion and coalescence processes in CoDy and CoHo alloys and Co/Gd(ultrathin) bilayers [KUN25a, KUN25b]. These studies extend the range of potential materials eligible for AO-HIS based technologies. Surprisingly, they did not observe this AO-HIS mechanism in CoTb alloys [KUN26]. They argued that this specific toggle switching, which occurs on the µs time scale, is achieved when the Co sublattice is fully demagnetized and the characteristic demagnetization times $\tau_{RE}$ of the RE (Gd, Dy or Ho) sublattices is longer than that of the Co which allows a flow of angular momentum from the RE to the Co sublattices and the angular momentum transfer from Co is reduced. This last situation occurs either by the use of high spin orbit coupling RE (Dy or Ho) [KUN25a] or low concentration of Gd [KUN25b].

In literature, characteristic demagnetization times of less than 200 fs were reported for Co in Co-RE (RE = Gd, Tb, Dy) alloys while the corresponding characteristic demagnetization times for the RE were longer ($\tau_{Gd} \sim 0.48$ ps, $\tau_{Tb} < 0.5$ ps and $\tau_{Dy} > 0.6$ ps at room temperature) [LOP13, BER14, RAD15, FER17, HEN19, FER21]. Based on the available data sets, Kunyangyuen et al. have claimed that $\tau_{Tb}$ was too short to observe AO-HIS and predicted that the characteristic demagnetization time of the Ho in CoHo alloys should be at least as long as that of the Dy. Unfortunately, no data has been reported so far for CoHo alloys that supports their reversal model. Furthermore, the reported simultaneous increase in spin–orbit coupling strength and demagnetization time appears counterintuitive, since stronger spin–orbit coupling is generally expected to promote spin-flip scattering and thus faster demagnetization [KOO10].

In this work, we have studied the laser induced ultrafast dynamics of Co 3d and Ho 4f spins in a $Co_{80}Ho_{20}$ ferrimagnetic alloy by means of time-resolved magneto-optical Kerr effect (TR-MOKE) magnetometry and time-resolved X-ray magnetic circular dichroism (TR-XMCD) spectroscopy [HOL14]. We observe a characteristic demagnetization time $\tau_{Ho} = 0.87 \pm 0.15$ ps for the Ho and $\tau_{Co} = 0.22 \pm 0.01$ ps for the Co sub-lattices at T = 300 K. These values are consistent with the scenario proposed by Kunyangyuen et al. [KUN25a] for the occurrence of domain motion AO-HIS in Co-RE alloys.

## 2. *Material and methods*

The measurements were performed on $[Ta(5)/Cu(20)]_3/Ta(5)/Co_{80}Ho_{20}(18)/Pt(3)/Al(3)$ (in nm) multilayers deposited by DC magnetron sputtering on a $Si_3N_4$ membrane. The alloy was capped with a Pt(3)/Al(3) bi-layer to prevent oxidation while the $[Ta(5)/Cu(20)]_3/Ta(5)$ multilayer is used as heat sink. The static magnetic properties of the CoHo alloy were characterized by mean of X-ray Magnetic Circular Dichroism spectroscopy with the ALICE reflectometer [ABR15] at the PM3 beamline at BESSYII synchrotron [KAC15], Helmholtz-Zentrum Berlin (HZB). The XAS spectra were acquired by measuring the transmission of circularly polarized X-ray as a function of the incident photon energy. The measurements were performed at normal incidence with respect to the X-ray beam. The XAS spectra were recorded for two opposite magnetic field directions (H±), with the field applied along the X-ray propagation axis, to extract the XMCD amplitude at the Co $L_3$ and Ho $M_5$ absorption edges.

Hysteresis loops were acquired by tuning the incoming photon energy at the Co $L_3$ absorption edge and monitoring the transmission as a function of the magnetic field. The TR-MOKE experiments were performed at room temperature in a polar configuration using a Light Conversion's PHAROS-SP system. A magnetic field was applied along the laser propagation axis. The laser repetition rate was set to 5 kHz, while the wavelengths for the pump and probe pulses were set to 1030 nm (fundamental) and 515 nm (frequency doubled by a BBO crystal), respectively. At this wavelength, TR-MOKE probes the Co sublattices in the Co-RE alloys. The TR-XMCD experiments were carried out in the Femtoslicing-Dynamax station at the UE56/1-ZPM beamline at the BESSY II Synchrotron radiation facility operated by the Helmholtz-Zentrum Berlin für Materialien und Energie [HOL14]. The magnetization dynamics have been measured by monitoring the transmission of 100 fs-circularly polarized X-ray pulses tuned to specific core level absorption edges as a function of a pump-probe delay for two opposite directions of the magnetic field [FER21]. The magnetic field, provided by the Dynamax magnet, was applied along the propagation axis of both the IR laser and the X-ray beam during the experiment. The photon energy was set to either the Co $L_3$ or Ho $M_5$ absorption edge using reflection zone plate monochromators on the UE56/1-ZPM instrument. The full width at half maximum (FWHM) of the 800nm pump laser was set to approximately 500 μm to ensure homogeneous pumping over the probed area of the sample (FWHM of X-ray pulse ~ 200 μm).

### 3. ***Experimental results and discussion***

The hysteresis loops recorded at the Co $L_3$ edge at 80 and 300 K show that the CoHo alloy displays out-of-plane magnetic anisotropy with coercive field $H_c$ (80K) = 57 mT and $H_c$ (300K) = 12 mT (figure 1). The sign inversion denotes the existence of a magnetic compensation temperature between 80 and 300K, confirming the nominal composition. The XAS spectra at the Co $L_3$ (figures 2a and 2b) and Ho $M_5$ (figures 2c and 2d) edges show that the XMCD amplitude remains nearly constant for the Co sublattice (figure 2e), while it decreases by a factor of approximately 1.8 for the Ho sublattice (figure 2f) when the temperature is increased from 80 to 300 K. Figures 2c and 2d also show that the multiplet $M_J$-state occupations change with temperature, following the Boltzmann distribution. Despite the redistribution of $M_J$-state occupations when the temperature is increased from 80 to 300 K, the shape of the XMCD spectra is not affected, only their amplitude changes. Such behaviors were also previously reported for CoDy alloys [AGU07, CHE15]. For the TR-XMCD experiments, the laser fluence was set at 8 mJ/cm² and the cryostat temperature was set to $T_{cryo}$ = 200 K to compensate the laser induced DC-heating ($\Delta T$). Therefore, the working temperature $T = T_{cryo} + \Delta T$ was close to room temperature. This matches the experimental conditions of Kunyangyuen et al., who investigated HI-AOS in Co-RE alloys at room temperature [KUN25a]. The Curie temperature ($T_{Curie}$) of $Co_{80}Ho_{20}$ alloys cannot be found in literature but interpolation of tabulated data by Hansen et al. suggest that it is larger than that of $Co_{80}Dy_{20}$ alloys ($T_{Curie} \sim 700$ K) [HAN91]. Therefore, under these conditions, we estimate that the temperature is at least 400 K below $T_{Curie}$ (thus $T^* > 400$ K, with $T^* = T_{Curie} - T$) [FER21]. The transient XMCD signal recorded at the Co $L_3$ and Ho $M_5$ edges are displayed in figure 3a. The TR-XMCD curves were fitted with exponential (decay and recovery) functions convolved with a Gaussian function which accounts for the experimental time resolution of 120 fs (solid lines in figure 3a). In figure 3b, TR-MOKE curve as well as the corresponding fits are shown. It is worth noticing that TR-MOKE experiments were carried out in ambient air and thus, we did not observe any significant signature of laser induced DC heating. On both TR-XMCD at Co $L_3$ and TR-MOKE traces, we observed a fast quenching of the Co 3d magnetization, with a demagnetization amplitude around 25 %, followed by a fast recovery. The experimental signal to noise ratio of the TR-XMCD at the Co $L_3$ edge results in large error bars, but from TR-MOKE we have extracted a characteristic demagnetization time of $\tau_{Co} = 0.22 \pm 0.01$ ps for the Co sub-lattice, which is identical to that previously reported for CoGd, CoTb and CoDy alloys [LOP13, BER14, FER17, FER21, ABR21]. It seems that the characteristic demagnetization time of the Co sub-lattices does not depend dramatically on the rare-earth elements. In contrast, we observe a longer demagnetization time for the Ho 4f sublattice compared to that of the Co, with a maximum demagnetization amplitude of about 60% reached after several ps. The larger demagnetization amplitude of the RE sublattice compared with that of the Co was also reported in CoDy alloys [FER17]. This effect is caused by the sharp reduction of the RE magnetization with increasing temperature. The Co magnetization has almost fully recovered, while the Ho sublattice is still demagnetizing as previously observed in CoDy alloys [RAD15, FER17, FER21, ABR21]. We have extracted a characteristic demagnetization time $\tau_{Ho} = 0.87 \pm 0.15$ ps for the Ho sub-lattice. This value is slightly longer than $\tau_{Dy} = 0.57 \pm 0.09$ ps, reported for the Dy sub-lattice in CoDy alloys at $T^* = 400$K, and substantially longer than the demagnetization times reported for the Tb sublattice in CoTb alloys ($\tau_{Tb} < 0.5$ ps) [LOP13, BER14]. This observation is consistent with the claim by Kunyangyuen et al. regarding the importance of the characteristic demagnetization times of both sublattices as a criterion for AO-HIS driven by domain motion.

## 4. *Conclusions*

We have investigated the laser induced ultrafast dynamics of Ho 4f spins in a ferrimagnetic CoHo alloy by element- and time-resolved XMCD. We have reported on characteristic demagnetization times $\tau_{Co} = 0.22 \pm 0.01$ ps and $\tau_{Ho} = 0.87 \pm 0.15$ ps for the Co and Ho sub-lattices, respectively. We have also observed that the Co magnetization has almost fully recovered by 3 ps while the Ho sublattice is still demagnetizing. Our experimental data are thus consistent with the claim of Kunyangyuen et al. that distinct Co 3d and RE 4f spin dynamics are required to enable domain motion driven AO-HIS in Co-RE alloys [KUN25a, KUN25b, KUN26]. In these CoRE alloys, the Tb 4f dynamics is faster than that of the Ho and Dy which may explain the absence of AO-HIS in CoTb alloys. Finally, this work adds to the scarce experimental reports on element- and time-resolved investigation on laser induced Co 3d and RE 4f dynamics in CoRE alloys [RAD11, LOP13, BER14, RAD15, FER17, HEN19, FER21, ABR21].

***Figures***:

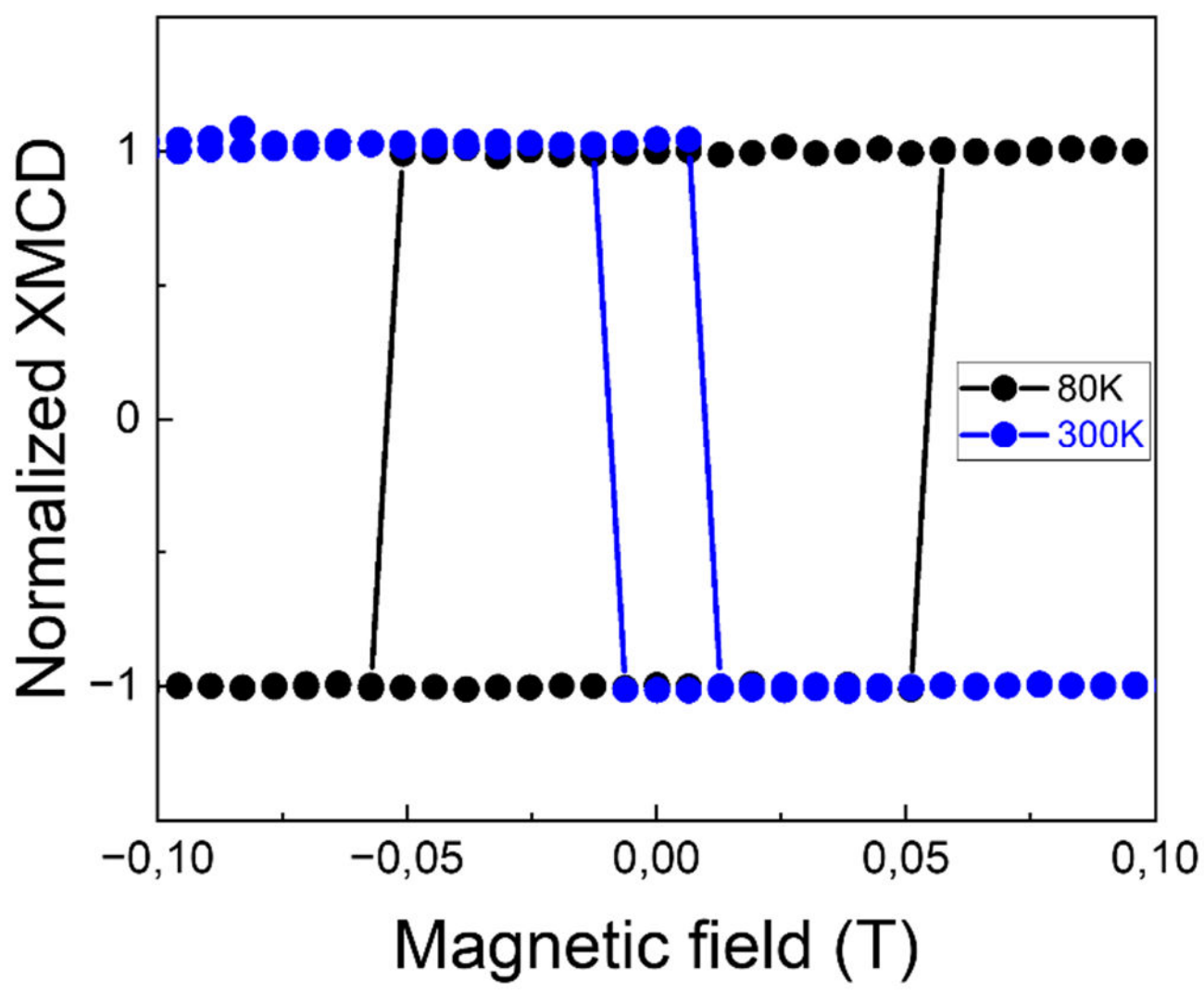

***Figure 1****: Hysteresis recorded at photon energy tuned at Co $L_3$ edge at T = 80 K (black curve) and 300 K (blue curve). Reversal in sign of hysteresis indicates that magnetic compensation temperature lies between 80 and 300 K.*

.

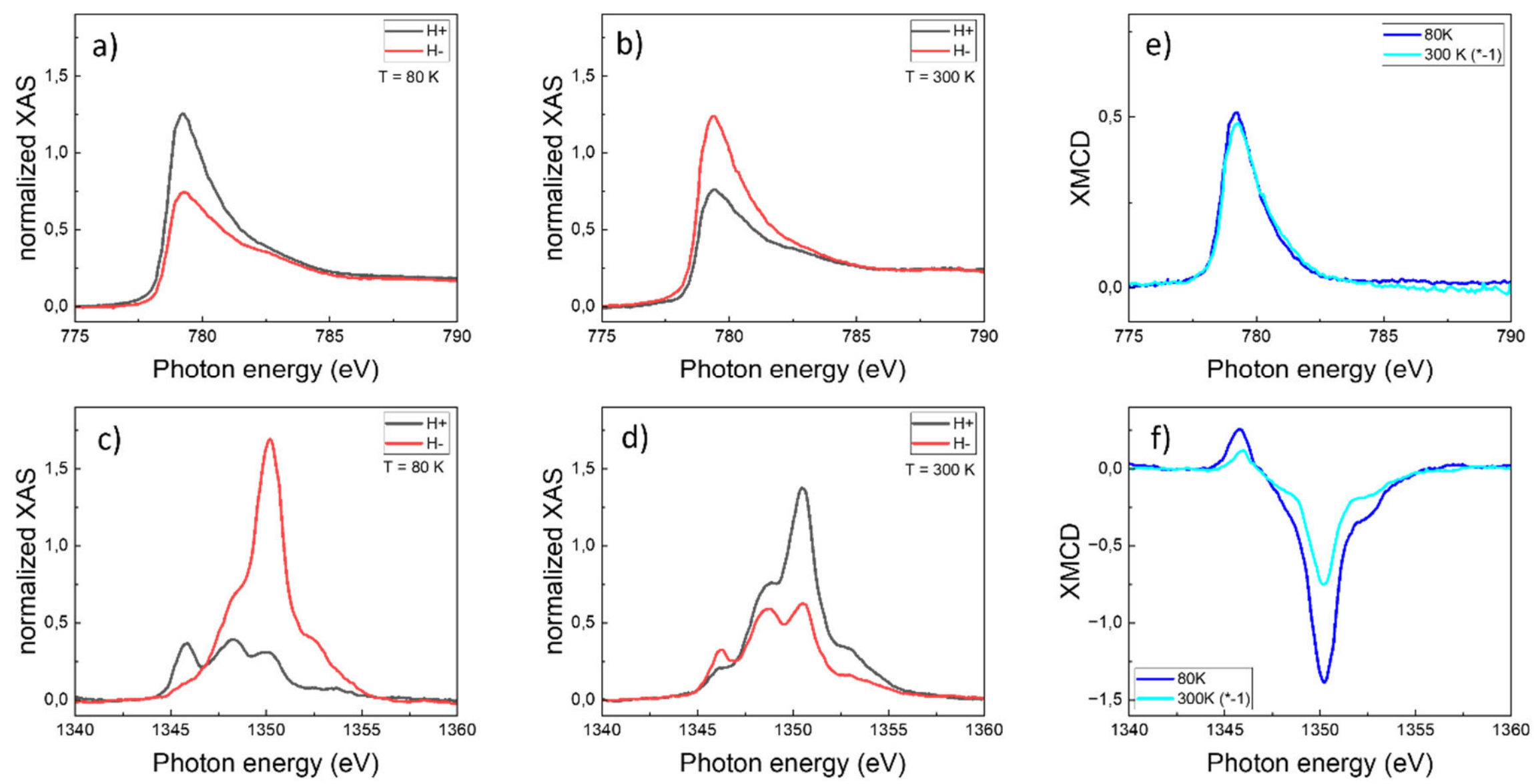

*Figure 2: Co $L_3$ (a, b) and Ho $M_5$ (c, d) XAS spectra recorded in transmission geometry as a function of incident photon energy at temperatures, T = 80 K and 300 K, respectively. The black ($H_+$) and red ($H_-$) curves correspond to opposite magnetic field directions. (e, f) Corresponding XMCD ($H_+$ - $H_-$) spectra obtained from the XAS spectra recorded for $H_+$ and $H_-$.*

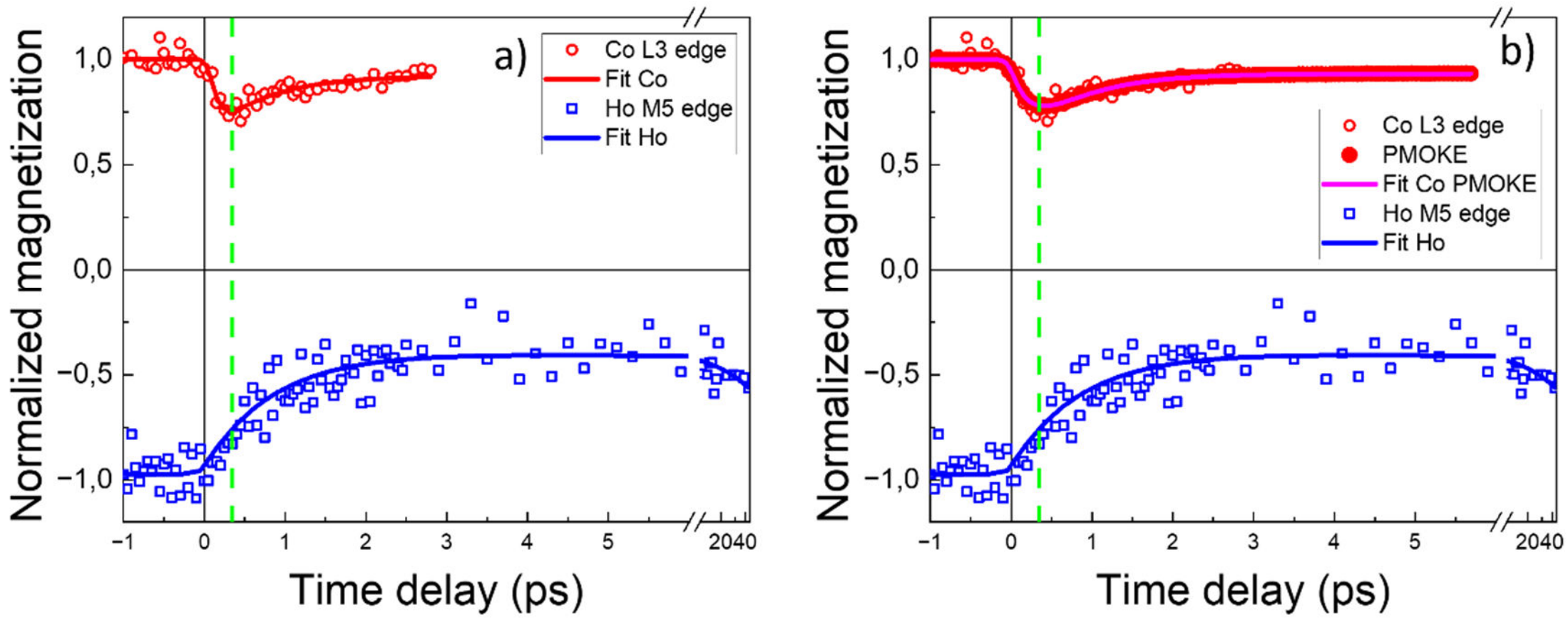

*Figure 3: Ultrafast demagnetization dynamics of the Co 3d (red open circles) and Ho 4f (blue open squares) sublattices in a CoHo alloy measured at Tcryo=200 K with a laser fluence of 8 mJ/cm2 using TR-XMCD. The solid red and blue lines are fits using an exponential function with a Gaussian convolution. (b) Demagnetization dynamics of the CoHo alloy measured by TR-MOKE. TR-MOKE primarily probes the Co sublattice in Co-RE alloys. Here, solid red circles, superimposed with Co L3 dynamics measured by TR-XMCD. Solid pink line is the fit. The vertical green dashed line marks the sharp distinction between demagnetization dynamics of the Co and Ho sublattices: the Co sublattice has reached its maximum demagnetization, whereas the Ho sublattice is demagnetized by only 25%.*

## ***Acknowledgments***:

We thank HZB for the allocation of synchrotron radiation beamtime. We thank Grégory Malinowski and Daniel Lacour for their help in this research topic. This work was supported by the Institute Carnot ICEEL, the Région Grand Est, the interdisciplinary project LUE "MAT-PULSE", part of the French PIA project "Lorraine Université d'Excellence" reference ANR-15-IDEX-04-LUE, a European Union Program. This work was supported by the ANR through the project ANR−21-CE42-0004-01 (EXPERTISE) and the France 2030 government grants PEPR SPIN (ANR-22-EXSP 0002), and PEPR SPIN –SPINMAT ANR-22-EXSP-0007. This work was supported by the Deutsche Forschungsgemeinschaft (DFG, German Research Foundation) - Project822No. 328545488 - TRR 227, Project A03.

The authors have no competing interests to declare.